# Improved understanding of the growth of blocky alpha in welded Zircaloy-4

Ruth M. Birch[1,2]*, T. Ben Britton[1,2]

1. Department of Materials, Imperial College London, Exhibition Road, London, UK, SW7 2AZ
2. Department of Materials Engineering, University of British Columbia, Frank Forward Building, 309-6350 Stores Road, Vancouver, BC, Canada V6T 1Z4

*corresponding author: ruth.birch@ubc.ca

## Abstract

Zirconium alloys are widely used in nuclear reactors as fuel cladding materials. Fuel cladding is used to contain the nuclear fuel and cladding tubes are typically sealed using welds. Welding of zirconium alloys can result in changes in the local microstructure, with the potential to grow so called 'blocky α' grains in the welded region during subsequent thermal processing and these blocky α grains have the potential to be detrimental to the integrity of the component. In this work, complimentary heating experiments with *ex situ* and *in situ* electron backscatter diffraction (EBSD) analysis are used to aid understanding of the blocky α grain growth within the weld region. These experiments reveal that blocky α grain growth is related to the parent β (high temperature) microstructure, as grains grow adjacent to a prior β grain boundary and the orientation of the growing α grain can be explained using neighbourhood orientations from this prior β grain. This growth mechanism is explained via a simple mechanism which is related to the α grain orientations, grain boundary structures and local stored energy. Ultimately, our findings indicate that the likely grain growth (size and morphology) across the weld region can now be predicted from the initial as-welded microstructure.

## Keywords

Zirconium, grain growth, welding, in situ, Electron backscattered diffraction (EBSD)

## 1 Introduction

Zircaloy-4 is used extensively in the nuclear industry as cladding for fuel rods and other structural components. To retain and optimise the desired material properties of zirconium-based components, e.g. neutronic performance and mechanical strength, the microstructure within a component is closely controlled during manufacture. Unfortunately, while bulk parts can have the grain structure and crystallographic texture controlled during process, a subsequent joining operation can result in local microstructural changes that may be of significant concern. For electron beam welds in Zircaloy-4, Taylor et al.[1] have shown that heat treatments of welded structures can result in abnormal grain growth. This significant microstructural transformation was localised within the welded region. The initial grain structure is replaced with very large so called 'blocky α' grains (>300 μm in size with irregular and wavy boundaries [2], [3]) which reduces the grain boundary density and changes the local crystallographic texture significantly and therefore has the potential to impact component performance.

Blocky α grains have been observed in Zr alloys after heat treating strained material [2]–[4], in weld microstructures [1], [5] and in β-quenched Zircaloy-4 [6]. The blocky α microstructure is significantly different to the microstructures present both before welding and adjacent to the weld, and blocky α formation is thought to be detrimental to performance of the components. This can be a safety concern for thin-walled components, where the resultant grain size can approach the wall thickness.

To understand the change in grain structure, it is important to recall that in Zr alloys there is a solid phase change above ~865 °C [7] from the low temperature α phase, which has a hexagonally close packed (HCP) microstructure, to the high temperature β phase, which has a body centred cubic (BCC) crystal structure. These phases have a well-defined orientation relationship known as Burgers' orientation relationship (BOR) [8]:

$$(0001)_{\alpha}//\{110\}_{\beta} \text{ and } \langle 11\bar{2}0\rangle_{\alpha}//\langle 1\bar{1}1\rangle_{\beta}$$

During electron beam (EB) welding fusion occurs due to local melting of the component due to the flux of electrons. In addition to this fusion zone, there is a region nearby called the 'heat affected zone'. In the fusion and heat affected zone, material can reach temperatures above the β transus, and therefore the initial α-grains will transform into the β phase and form much larger β-grains. Once the electron beam is moved away from the region, these grains will rapidly cool, and the β-phase will rapidly cool and undergo a solid-state transformation into α-phase. This means that the final microstructure of regions that had transformed into the β-phase is related to the structure of the prior-β microstructure.

Transformation from β phase to α phase is a solid-state transformation and so the high temperature β phase microstructure can be reconstructed with local crystallographic analysis of the low temperature α phase grains, using the BOR and electron backscatter diffraction (EBSD) based microstructural analysis [9], [10], [19], [11]–[18]. These reconstruction methods typically rely on analysis of the local orientation and/or misorientations of points and/or grains in EBSD maps, together with the crystallography of the BOR, to reconstruct the parent microstructure and a brief search of the literature reveals that there are a number of automated codes are available for parent grain reconstruction in Zr alloys, both proprietary and free. In this work we use ParentBOR [20] which is available on GitHub (https://github.com/ExpMicroMech/ParentBOR).

The formation of blocky α remains an important and outstanding question for the manufacture of zirconium-alloy cladding materials, and the potential to use EBSD to uncover mechanisms that help us predict the dramatic nature of this microstructure change in welded zirconium-alloy parts motivates the present manuscript. In this work, we use a combined *ex situ* and *in situ* heating experiment to aid understanding of the grain growth mechanisms to assist in the prediction of the final microstructures formed. This experimental approach enables us to propose a mechanism that can explain the onset of growth of blocky α and its relationship to the prior β grain boundaries.

## 2 Experimental

To investigate blocky α grain growth in electron beam welds, two complimentary *ex situ* and an *in situ* experiments were carried out on electron beam welded samples.

## 2.1 Samples

Zircaloy-4 electron beam weld samples were provided by Rolls Royce plc. These samples were manufactured using a high energy density, autogenous electron beam weld to join a thin, flexible member and a thick component together, where the thick component provided a large heat sink effect to achieve to fast cooling of the weld and relatively high residual stress in the weld.

For the experiments performed here, the welded samples were then sectioned into slices approx. 1.5 mm thick using a Buehler Isomet 1000 slow saw (gravity fed). All samples were ground using 800-2000 grit SiC papers followed by a 4.5 hr polish with OP-S. Prior to any scanning electron microscopy (SEM) analysis, the samples were broad ion beam polished (BIB) [21] using a Gatan PECSII machine for 15 mins at 8 keV, 8° tilt, 1 rpm, dual modulation. BIB was also used to remove any oxide build up after the *in situ* experiment.

## *2.2 Ex situ*

The *ex situ* experiment was designed to look (macroscopically) at where within the weld region the grains grow and whether the microstructure was related to the prior β microstructure. In this case, a sample was polished, imaged, then encapsulated in a quartz tube, backfilled with Ar, and heated to 760 °C (below β transus) for 4 hrs in a Lenton box furnace, followed by air cooling. The heating duration was chosen to have grain growth across the entire weld region.

## *2.3 In situ*

The *in situ* experiment was designed to look at the onset of grain growth and used an interrupted heating cycle, with 10 min holds at temperature (800 °C on the Murano thermocouple, >760 °C surface temperature) interspersed with holds at 500 °C (Murano thermocouple) to enable forescatter imaging of the surface (not discussed in this paper). The *in situ* heating experiment was carried out using a Gatan Murano heating stage in an FEI Quanta FEG 650 SEM. EBSD data was captured using the attached Bruker eFlashHD2 camera with a high temperature (Al coated) phosphor.

## 2.4 Characterization

In both cases, the microstructure was captured using forescatter imaging (i.e. electron channelling contrast imaging, ECCI) and EBSD at the start and end (and at an intermediate step in one case) to evaluate grain growth. EBSD data was processed using Bruker ESPRIT software and data exported to .h5 prior to analysis in MATLAB (R2018b) using MTEX version 5.4. EBSD patterns were captured using a 20 kV beam with a ~10 nA probe current at a working distance of ~15 mm for *ex situ* experiments and ~20 mm for *in situ* experiments, combined with 70° sample tilt. The diffraction patterns were captured at a resolution of 160 x 120 pixels using a 7 ms exposure time and 1 µm step size. Indexing was performed online within ESPRIT. The parent β microstructure was then reconstructed using previously developed ParentBOR code [20] which is available on GitHub. Grains were identified using a threshold angle of 4° and minimum grain size of 15 pixels. ParentBOR reconstruction settings in all cases were cut-off 4° and inflation power 1.4.

## 3 Results

The experimental results are reported first by exploring the *ex situ* experimental observations to understand the relative grain structure before and after heat treatments, and identify different regions at the end. This is followed by analysis from the *in situ* experiments, where the evolution of the grain structure is tracked to form a predictive model of the grain growth process.

### 3.1 *Ex situ* experimental observations

#### *3.1.1 Grain growth across the weld region*

Grain growth is seen across the entire weld region, as shown in Figure 1 where a series of EBSD maps is presented from regions of the same sample pre and post heat treatment. The regions were selected to survey the microstructure that includes bulk microstructure from the thin sample, through the weld, and into the second (larger) piece of zircaloy.

There are six regions that can be identified within the samples:

(1) The bulk sample above the weld. This region shows very little change before and after the heat treatment. Quantitative analysis of the microstructure here shows that the α-grain size is unchanged. Near the interface with the domain there is potentially growth of blocky α into this region. This upper region is originally from the thin plate sample.

(2) The heat affected zone and fusion zone above the weld line, which originally belonged to the thin plate sample. This region shows significant blocky α growth in the post heat treatment sample with elongation of α-grains away from the weld line and towards the upper region. The shape of these grains and their location indicates that they have grown upwards from the weld region.

(3) The weld centreline. This region is marked with α-grains that run left to right in the pre-heat treatment IPF maps and the region shown as a thin decorating line between large blocky α grains in the post-heat treatment IPF map.

(4) The fusion zone below the weld centreline has a distinct crystallographic texture (indicated by the green and blue grains (i.e. <a>/<a1+a2> along Z textures) which differ to the grains in the region below. Once these grains have been identified within the post-heat treatment map, it is possible to correlate these grains with a difference in grain orientation in the pre-heat treatment map, although the difference is more subtle.

(5) The heat affected zone below the weld. In the post heat treatment map, the grains are blocky α with grain orientations that significantly vary in crystal orientation from both the bulk material below and the fusion zone above.

(6) The bulk microstructure below, which (like the bulk above) shows very little change in microstructure before and after the heat treatment.

Identification of these regions highlights that within the weld region, the heat affected zone (HAZ) and fusion zone behave differently. In the HAZ there are relatively large regions with containing fine grains with a limited range of orientations. Grain growth in the HAZ then results in large blocky α grains related to the initial regions of similarity, albeit subdivided, as will be discussed. In the fusion zone, the very fine microstructure can be grouped into elongated regions of similar orientations that transform to elongated blocky α grains similar in size to the initial regions. Due to the weld geometry, it is easier to delineate between these regions in the lower half of the weld, but the same trends are seen.

Across the weld region many small grains remain, this is likely to be a result of competitive growth and gives an indication of the growth mechanism.

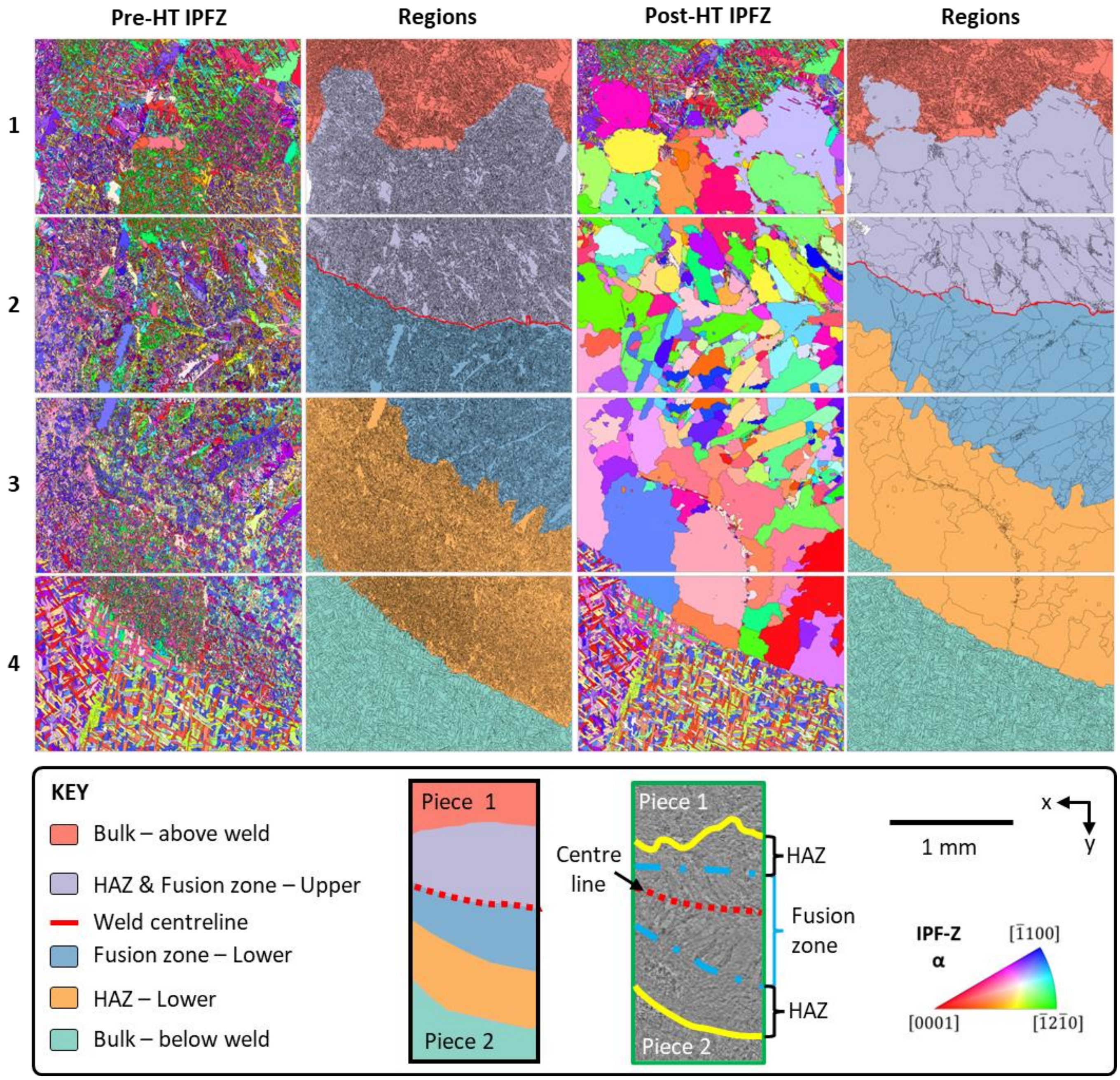

*Figure 1 – EBSD maps of the ex situ sample, with regions labelled. Pre-HT and post-HT data displayed using IPFZ colouring (columns 1 & 3) and the same maps coloured by region (columns 2 & 4). Regions: (i) bulk – above weld; (ii) upper half of weld (fusion zone & HAZ); (iii) weld centreline; (iv) lower half of weld – central region; (v) lower half of weld – HAZ; (vi) bulk – below weld.*

### *3.1.2 Relationship to prior β microstructure*

Understanding of the grain growth mechanism is unlocked when prior-β grain reconstruction is used in combination with the *ex situ* heating data. Figure 2a and b present this data in more detail and include pattern quality, crystal orientation as represented in the IPF-X and IPF-Z colour keys, and the grain boundary networks. Figure 2a shows the α microstructure; and 2b shows the results of the prior-β reconstruction as applied to this data.

Comparison of these maps shows that the blocky α is linked to the prior-β structure, but the blocky α structure is not formed from an α-variant within a prior-β growing to consume the existing prior-β grain. This can be observed not only from the segmentation of the prior-β structure after heat treatments, but also the change in crystallographic orientations of the reconstructed prior-β grains.

Combining the as measured α-maps and the reconstructed β-maps (Figure 2b) enable the extent of the regions within the weld structure to be more easily observed:

(i) In the above weld region, there are large equiaxed β in the bulk and these stay the same before and after HT.

(ii) The upper HAZ and fusion zone is now separated in the reconstruction, with elongated grains extending from the bulk region towards the weld centreline presumably as the material solidified and transformed last at/or near the weld centreline. At the top edge of the upper HAZ, some partial prior-β grains are retained which aids investigation of how the grains grow – these are discussed in section 3.1.3.

(iii) The weld centreline is decorated with a necklace of small grains pre-heat treatment, and this necklace is removed after the heat treatment but is left as the interface between the upper and lower HAZ/fusion zones.

(iv) The lower fusion zone shows similar elongation of the reconstructed β-grains again extending towards the weld centreline.

(v) The lower HAZ contains large reconstructed β-grains in the pre-HT maps which are subsequently subdivided into smaller β-grains in the post HT map.

(vi) The below weld bulk region remains the same before and after HT.

In the weld region (i.e. all areas excluding the bulk material above and below), there are many island grains present in the post HT maps. These island grains are similar in location and morphology to prior island grains seen in blocky α by Tong and Britton [3].

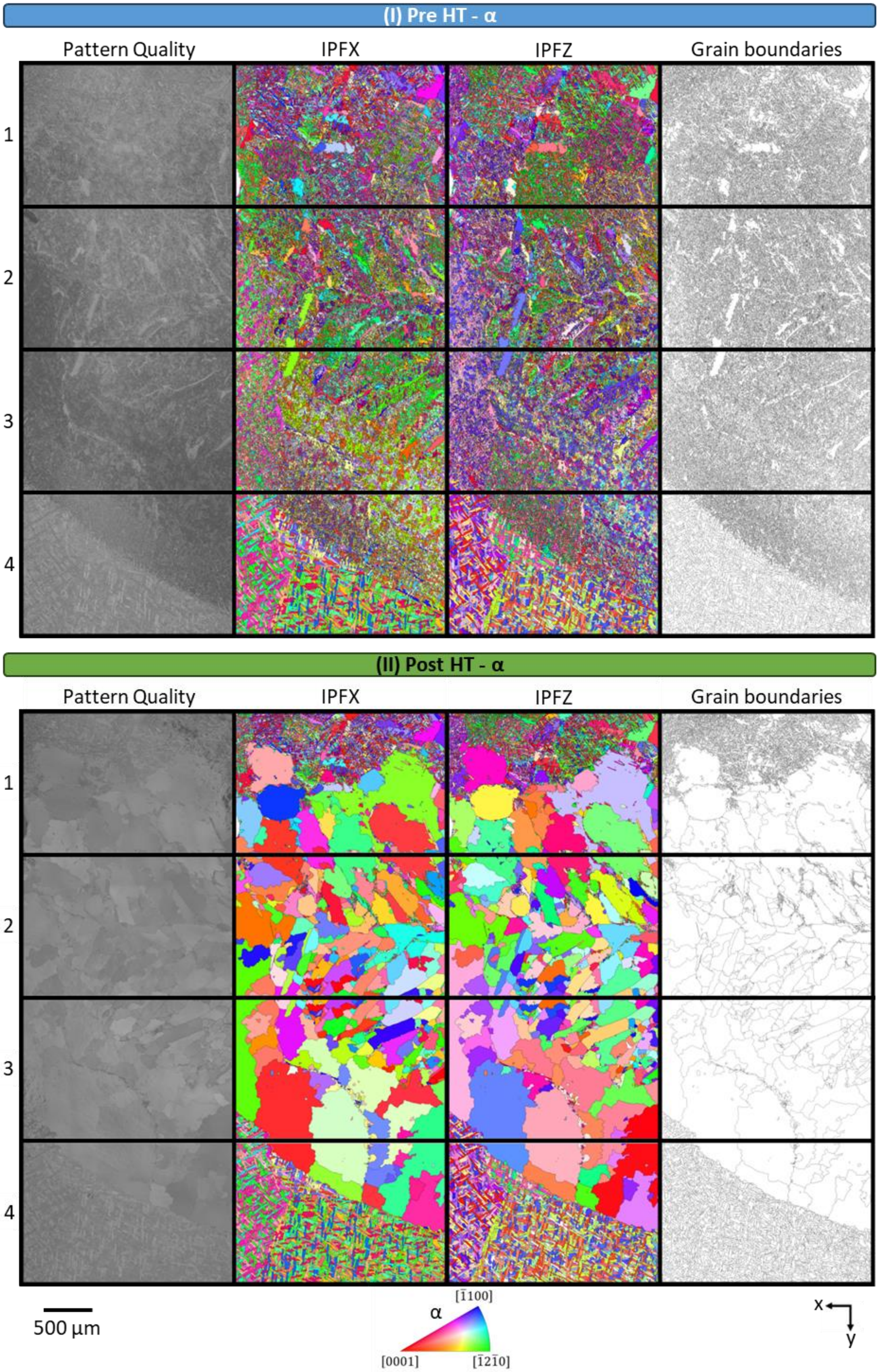

*Figure 2a - Ex situ EBSD results – α. Maps of pattern quality, crystal orientation (IPFX, IPFZ coloured) and grain boundary outline map (threshold > 4°) for (I) Pre-HT and (II) Post-HT.*

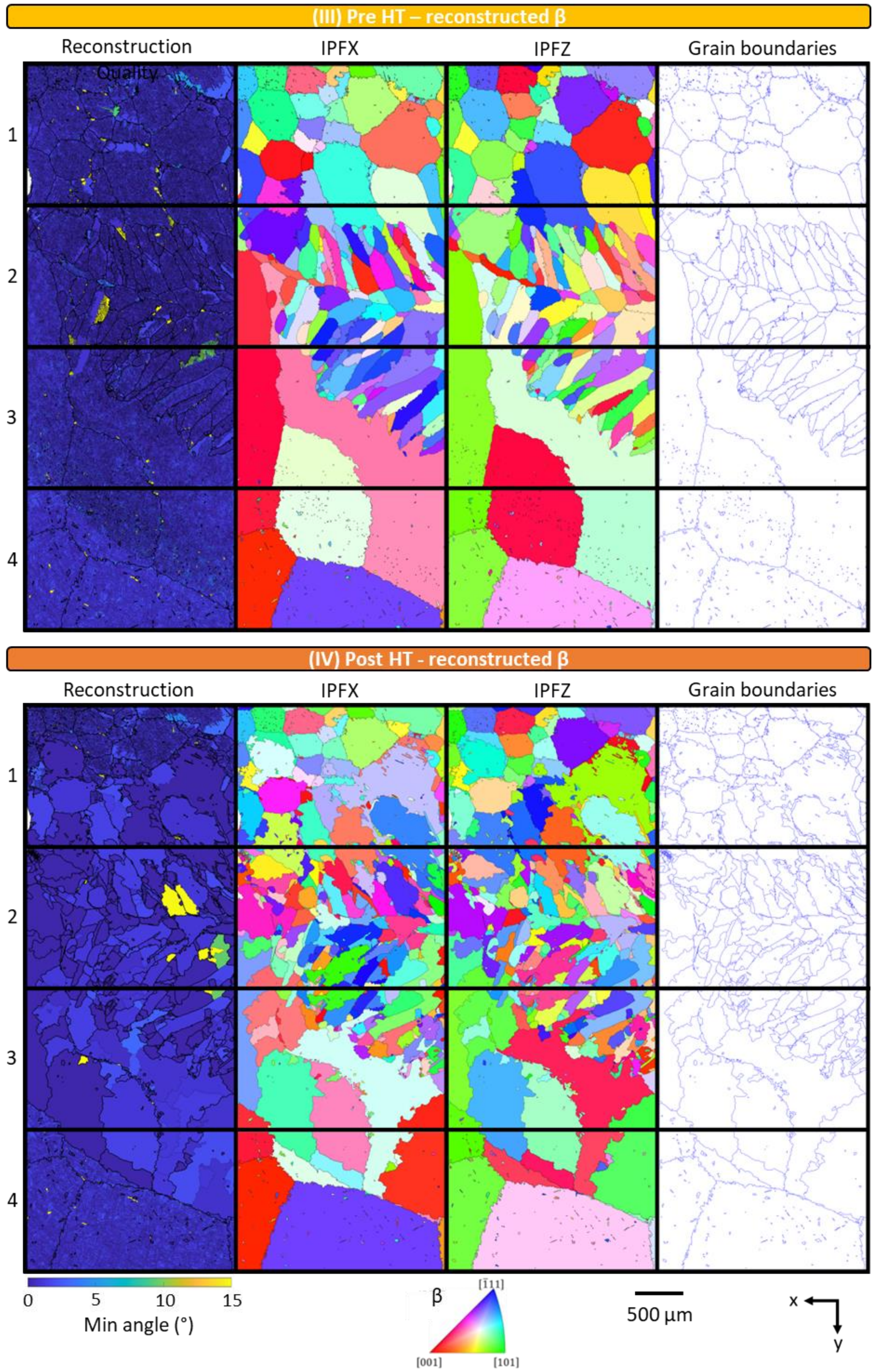

*Figure 2b - Ex situ EBSD results – reconstructed β. Maps of reconstruction quality, crystal orientation (IPFX, IPFZ coloured) and grain boundary outline map (threshold > 4°) for (III) Pre-HT and (IV) Post-HT.*

### 3.1.3 *Neighbourhood prior β orientations*

To elucidate a mechanism, we extract one example of grain growth as taken from the upper heat affected zone and explore the crystallographic relationships in detail within Figure 3. In this area, there is retention of the parent orientations after the heat treatment, but there is also one large blocky α grain that grows within this area. The post-HT grown blocky α extends across two parent grains from the pre-HT prior β reconstruction. Analysis of the pole figure data (see supplementary data) confirms that the α grain that grows ($\alpha_1$ – pink) was not present in the original region for either of the two parent β grain regions.

If we then look at the grains parent β domains surrounding this region using pole figure analysis, as in the lower section of Figure 3, there is good correlation with the α orientations from two of the surrounding β domains, indicating that this grain may be related to or have grown from one of these surrounding β domains. One of these cases ($\beta_5$) is illustrated in Figure 3, with the seed orientation at the grain boundary shown.

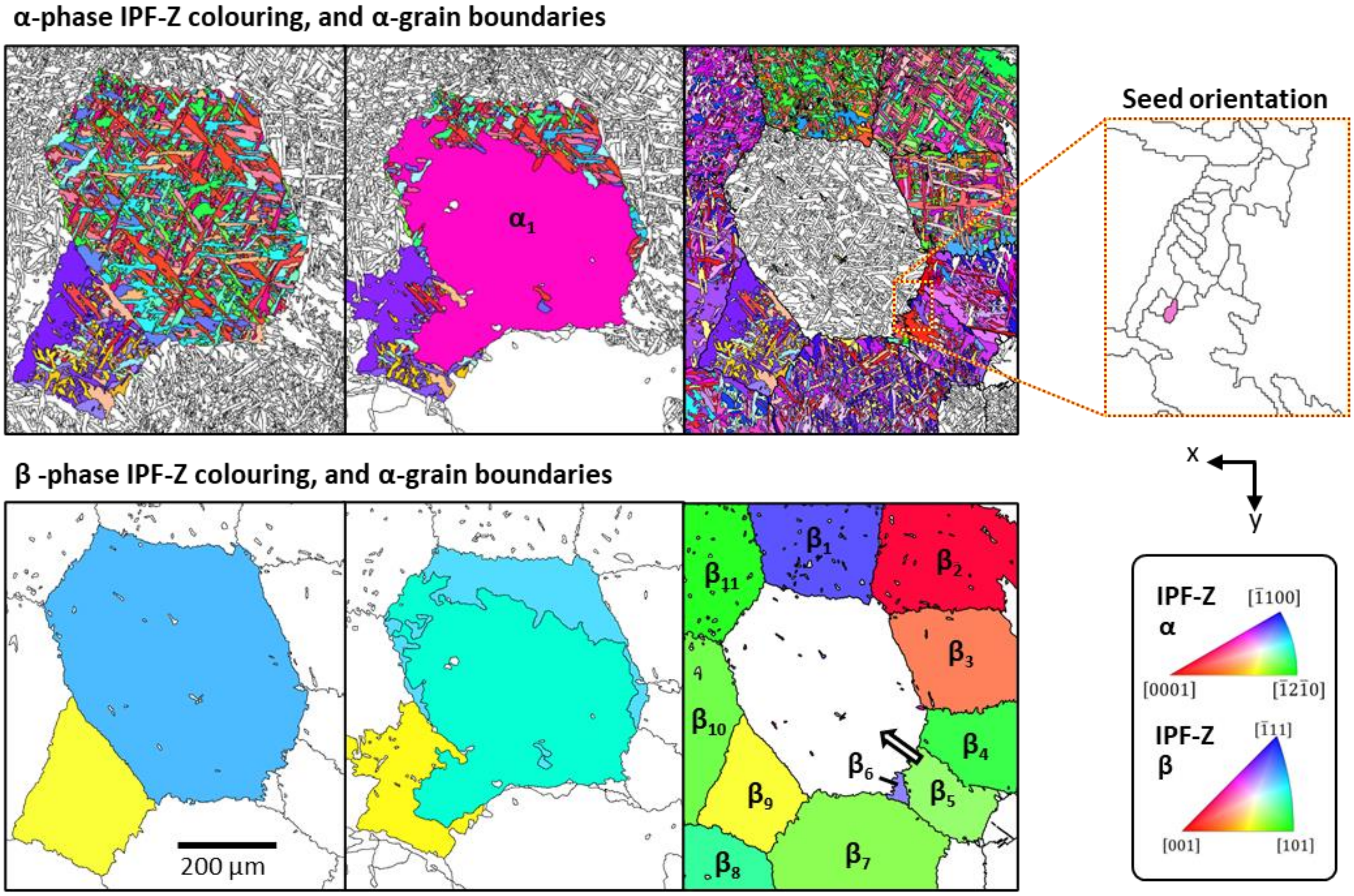

*Figure 3 – Parent β vs. post-HT orientation testing. Orientation maps (IPFZ) for pre-HT and post-HT α are shown in the top row, with the reconstructed β phase maps shown below. The arrow indicates the direction of growth for $\alpha_1$ from its origin within β grain 5. The seed orientation for the blocky α grain ($\alpha_1$) that grows is shown in the box. A more detailed version of this figure including full pole figure analysis can be found in the supplementary data.*

## 3.2 *In situ* experimental observations

The *in situ* experiment allows us to look at the initial onset of the grain growth and evaluate this further to develop a mechanism for grain growth.

### 3.2.1 *Initial onset of grain growth*

Figure 4 shows the initial and final microstructure for one of the *in situ* experiments. The region selected for *in situ* mapping is similar to the region shown in row 2 for the *ex situ* experiments (Figures 1-3) and this region contains the weld centreline.

The *in situ* observation of the changes reveals multiple areas of α-grain growth within this mapped region, and these regions of grain growth are correlated to the grain boundaries between parent-β grains within the reconstructed microstructures. These maps show that: (1) growth appears to originate from one parent grain and grow outwards into an adjacent parent grain in all cases (2) the grain growth is always related to the prior β grain boundaries (i.e. there is no grain growth inside the parent grains).

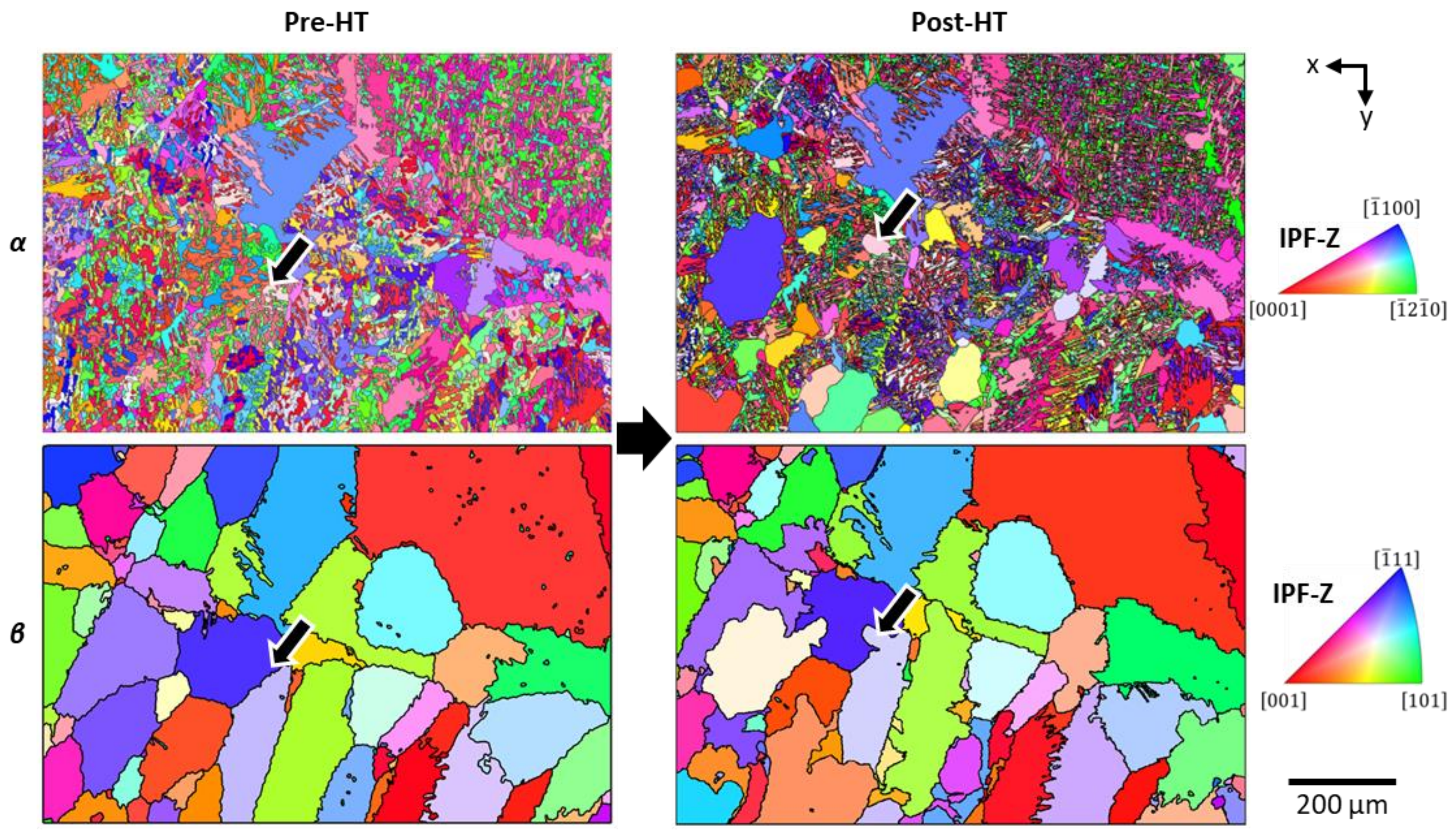

*Figure 4 – In situ evolution of grain growth, highlighting how one variant from a neighbour parent β grain grows inwards to form the new blocky α grain. Top row: α orientation map for pre and post HT; bottom row: reconstructed β orientation maps pre and post HT. The annotation arrows indicate an area where growth occurs (similar point on all maps) that is analysed further in the next section.*

Taking on example from Figure 4 (marked by the arrow), we can evaluate these observations by looking at the orientations in more detail in Figure 5. Here grain $\alpha_1$ appears to grow from the parent grain $\beta_2$ into $\beta_1$, across the parent (β/β) grain boundary. This is confirmed by the pole figure analysis (Figure 5 B) where the $\alpha_1$ orientation matches the BOR for $\beta_2$ but not $\beta_1$, indicating that the grown α region originated from $\beta_2$. Further analysis of the calculated α variants for the two β domain orientations (not shown) confirms this - we see that the α grain orientation correlates with one of the potential variants for $\beta_2$. Note that the α variants for the two β grains do not share any orientations, and this means that the grown α grain must have grown into the neighbouring β grain.

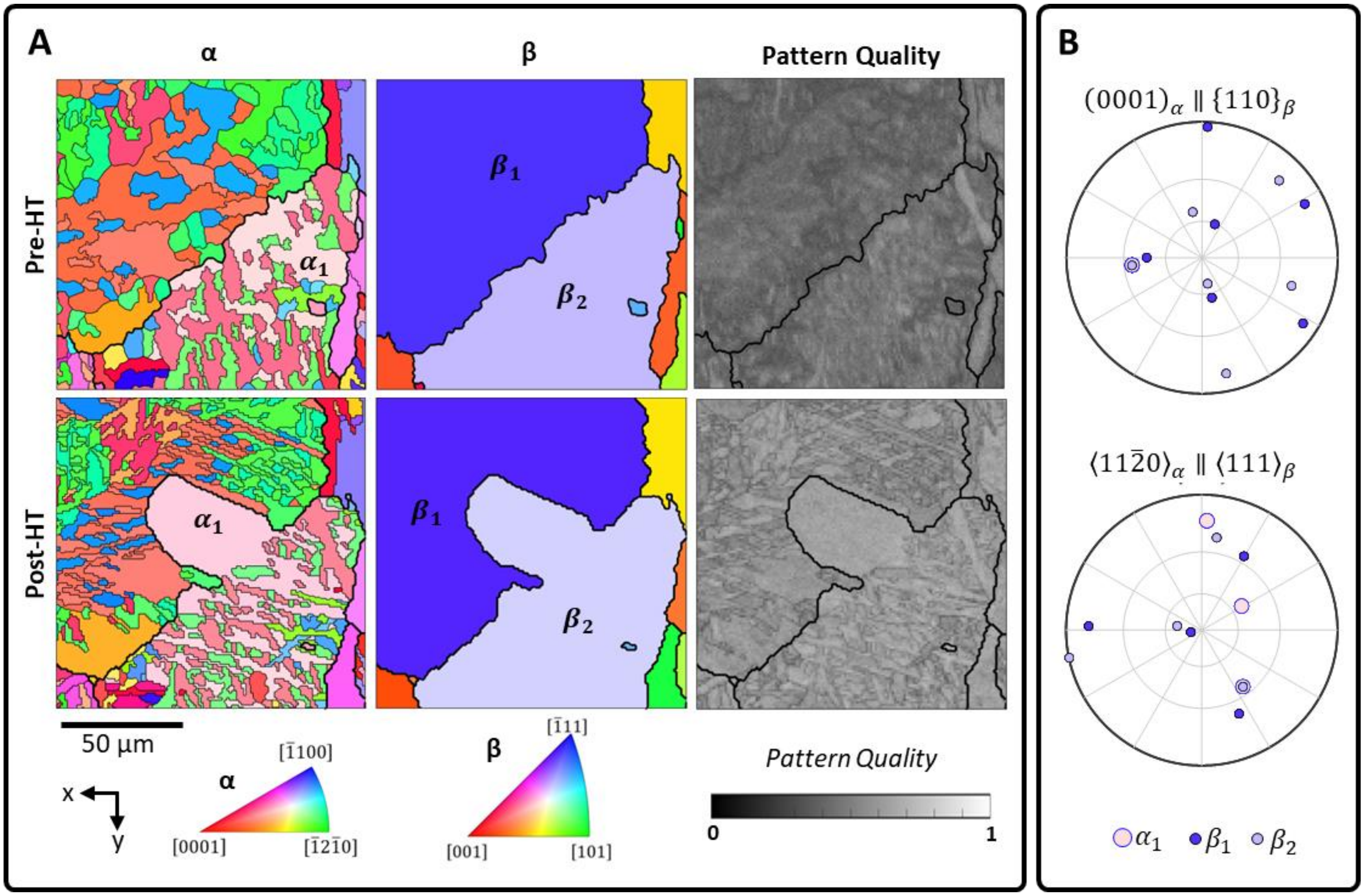

*Figure 5 – Example of the onset of blocky α grain growth - $\alpha_1$ grows from β domain $\beta_2$ into $\beta_1$ across the grain boundary. (A) pre-HT and post-HT maps of α orientations, reconstructed β orientations and pattern quality; (B) Pole figures for $\alpha_1$ showing overlap (and correlation with) with $\beta_2$ but not $\beta_1$.*

Further observations we can make from Figure 5 are that pattern quality improves locally when the grain grows, and the grain boundary curvature is high (higher stored energy) in the region that the grain grows into. Low pattern quality is often correlated with strain gradients within the interaction volume, i.e. high stored energy. Together these observations indicate that the new grains grow into regions of higher energy. This could be a driving force for grain growth which is consistent with the nucleation and growth mechanism of grain growth for blocky α proposed by Tong et al. [3]. To recap from this literature, Tong et al. proposed that blocky α grain growth was via strain induced grain boundary migration (SIBM), where the grains grow from regions of low strain energy into high strain energy regions.

# 4 Discussion

## 4.1 Predicting blocky α grain growth across the weld region

Observations of blocky α grain growth seen aligns well with prior work by Lusby [22] and Taylor et al. [1]. Notably, the blocky α grain growth affects the entire weld region after holding at 760° for 4 hr (i.e. ~100° sub-transus), with an observable distinction in the grain structure between the weld region and the bulk material.

Looking at the different regions within the weld area between the pre- and post-HT microstructures:

- **Bulk -** The α lath microstructure within the bulk material does not change significantly in terms of orientation or morphology.
- **HAZ -** The HAZ contains the largest blocky α grain growth across the weld region. The cooling rate for the welding affects the size of the HAZ and this in turn impacts the maximum size of the blocky α grains, as seen by the (generally) smaller grains in the HAZ of the top half of the weld. There are also smaller grains that are retained after grain growth, for example, the trail or necklace of smaller grains along a prior-β grain boundary in the lower half of the weld. These smaller grains generally have different orientations to the blocky α in this region. This indicates that there is competitive growth with certain orientations more likely to grow.
- **Fusion zone** - The fusion zone morphology most closely resembles the reconstructed β microstructure from the pre-HT map. The post-HT fusion zone microstructure is characterised by elongated blocky α growth starting from the centreline. Again, smaller α grains that have not grown are also contained within the region.

The grain growth patterns in the different regions means that the likely grain growth pattern can be predicted from the pre-HT microstructure. Practically this means that if you can control the weld microstructure (for example, minimising the HAZ size via different cooling rates), it is likely that you can control the post weld heat treatment microstructure.

## 4.2 Relationship to prior β microstructure

Comparing the reconstructed β microstructure from the pre-HT maps to the post-HT α microstructure, we observe that the post-HT microstructure is related to the prior-β microstructure. However, the prior-β microstructure is not recreated exactly in the post heat treatment microstructure (neither with the blocky α grains, nor the post HT reconstructed β microstructure), as subdivision and/or growth into adjacent prior-β regions occurs. This is found consistently across the entire map, though more noticeable in certain regions, for example, subdivision of the prior β domains is clearly observable in the HAZ (see Figure 2b).

The blocky α grain that grows in a prior-β spatial area does not originate from one of the (pre-HT) α grains within this prior-β grain (See section 3.1.3). Using orientation analysis, it is seen that blocky α grains grow from a neighbouring prior-β grain, so the growth occurs outwards from one prior β domain into another.

## 4.3 Blocky α grain growth

Blocky α grows within the weld region after the sub-transus heat treatment. The largest blocky α grains are located within the heat affected zone and these are also typically more equiaxed. Smaller and elongated blocky α grains are found to extend from the ‘inner’ edge of the heat affected zone towards the weld centreline. These grains have a similar morphology and direction to the blocky α grains in Tong and Britton [3], and echo the grain structure of the pre-HT β-reconstruction (i.e. prior-β) microstructure. This indicates that blocky α grows from nucleation sites and extends along regions of high residual strain (due to thermal expansion mismatch) and the associated templating of the blocky α grain growth with respect to the prior-β grain growth.

In all observed cases, grain growth is related to a prior-β grain boundary – no evidence was found of grains growing inside prior-β domains. These α grains grow outwards from one prior-β domain into an adjacent β domain and the orientation of the growing α grains can be explained using the neighbourhood orientations.

Grains grow across the original prior-β grain boundary from regions of low stored energy to regions of higher strain energy, as evidenced by the pattern quality maps, where darker regions tend to be related to higher stored energy. This is also evidenced by the short, curvy α grain boundaries within these darker areas that are higher energy than the adjacent areas with long/straight boundaries. Grain growth driven by stored energy is consistent with SIBM as a nucleation and growth mechanism, as proposed by Tong and Britton [3].

As the grains grow in the new domain, factors affecting the grain growth include pinning by 'cross' points of α laths in the microstructure and low grain boundary curvature. The cross points probably contain an SPP (ref) at the intersection of the α variants, and the lamella α extending from this seem to inhibit further blocky α growth. Grains that have long straight grain boundaries (as these often do) are lower energy and stable compared to the surrounding area and are less favourable for growth.

### 4.4 Factors affecting grain growth

As the grains grow in the new parent domain, growth appears to continue to be into regions of higher local strain, as indicated by the darker regions of pattern quality and the shorter curvier grain boundaries. There are two features which limit the grain growth – 'plus' shaped regions where several α grains meet, and long straight grain boundaries.

Within the prior-β regions, there is the occurrence of crosses of α grains, forming an 'x' motif which may be 'plus' shaped or 'star' shaped depending on the orientation of the parent β grain for the region and the sectioned plane. Looking at these in pattern quality, in some cases, the presence of a potential SPPs can be observed in the centre of these points (indicated by a darker region of pattern quality, although these points do not index so may have been removed during the polishing process). This would fit with observations by Holt [23], where SPPs act as nucleation points (depending on material composition) for some α grains during initial lamella α formation. Further analysis of these points is included in the supplementary data.

Growth often slows and the grain grows along (i.e. the grain boundary is retained) long straight boundaries within the microstructure, which are often connected to these 'cross' points. These grains will often have the long axis along {110} planes within the parent β grain for the region. These boundaries are more stable and lower energy than the shorter curvier boundaries where the grain grows initially, again supporting the argument that the grain grows into areas of high stored energy.

### 4.5 Grain growth model

Reconstructing the parent β microstructure facilitates the identification of the individual α variants (within each β domain). There are 12 unique α variants, which can be grouped into 4 groups with a shared <a> type direction or 6 pairs that share a (0001) plane (for more details see [20]). We observe that with the β domains there are distinct grouping of α grains that share an <a> type direction, with pairs with shared (0001) planes throughout (see supplementary data or Figure 6 insert).

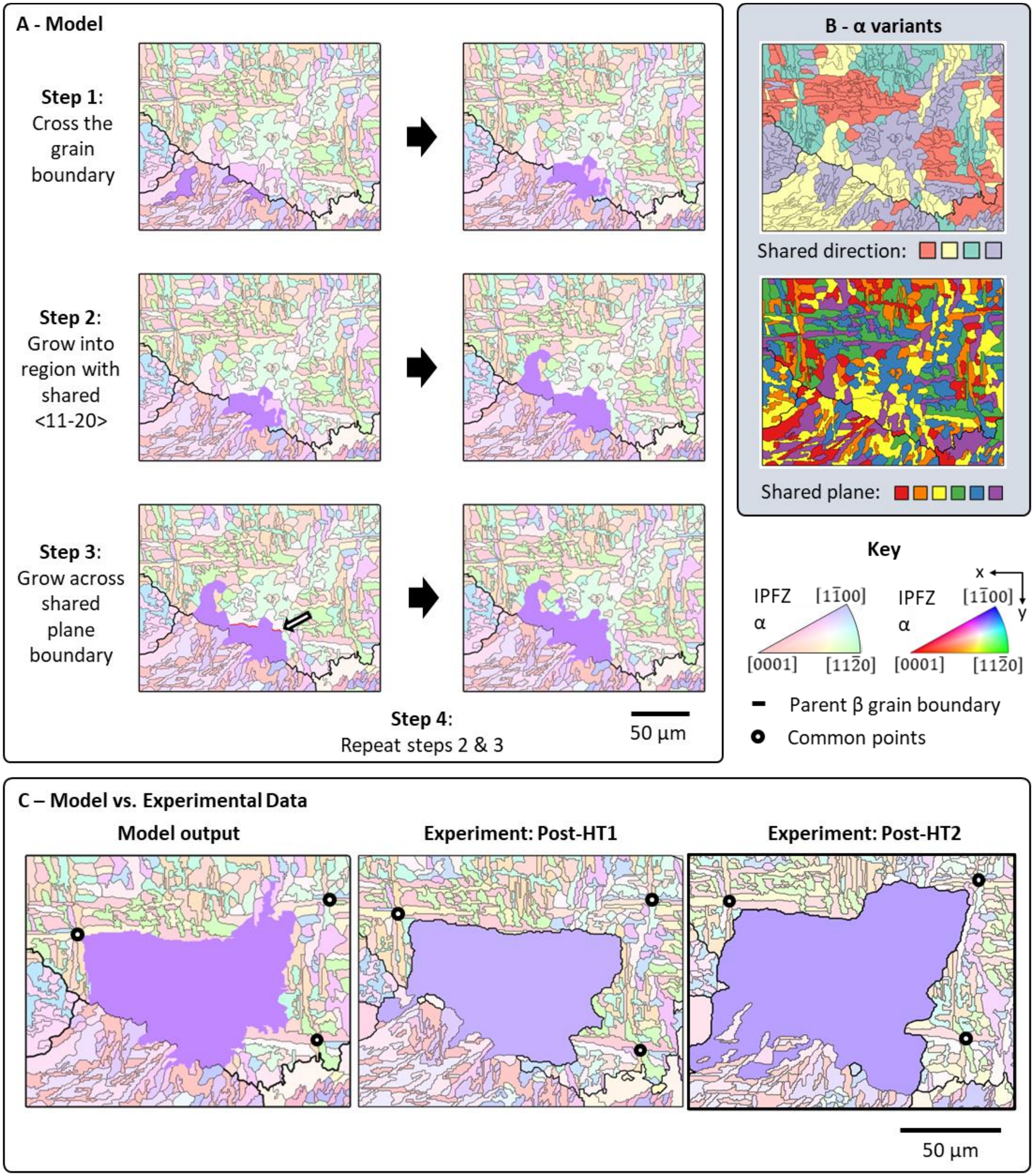

*Figure 6 - Proposed grain growth hypothesis. (A) Steps for model; (B) α variant maps coloured by shared direction (pastel colours) or plane (bright colours) for reference; (C) Model output compared with the output after the first (HT1) and second (HT2) heat treatments. Common points across the maps are marked by the white and black circles.*

A grain growth hypothesis is proposed using the shared planes and directions between the α variants as a pathway of easy growth within the new β domain (see Figure 6): (Step 1) Cross the boundary - α grain grows across the original β grain boundary location into the new prior-β domain. (Step 2) Grow to fill shared $\langle 11\bar{2}0 \rangle$ region - In the region where the α grain is, growth occurs between α variants with a shared $\langle 11\bar{2}0 \rangle$ direction. (Step 3) Grow across shared $(0001)$ plane grain boundary - Growth occurs across grain boundary where the adjacent grain shares a common $(0001)$ plane. (Step 4) Repeat steps 2 & 3 - growth continues via groups of α variants with shared $\langle 11\bar{2}0 \rangle$ directions and moves between these groups via α grain pairs with shared $(0001)$ planes.

This was tested on an *in situ* sample that had two heat treatments, so we see an intermediate growth stage. These maps are presented in comparison with the model in Figure 6 C. We see good agreement with the model in both cases, with the longer heat treatment resembling the model output better. Note that in this case, the known pinning points were taken into account.

## 5 Conclusions

Complementary *ex situ* and *in situ* experiments were carried out to investigate blocky α grain growth in electron beam welded Zircaloy-4. This work has revealed the following key findings are:

- Grain growth occurs within the weld region only and there is a clear transition to the original fine grain microstructure even after blocky α grain growth occurs. This aligns with prior work that used a bead-on-plate electron beam weld [1].
- Blocky grains grow outwards from one prior β domain into another from one prior β domain into a neighbour. This growth is by strain induced boundary migration (SIBM), as evidenced by the indications of higher stored energy (poor pattern quality and curved grain boundary) on the side of the boundary into which the nucleating grain grows.
- Factors affecting grain growth within grains include: pinning by 'cross' points of α laths in the microstructure, where the cross points probably contain an SPP at the intersection of the α variants, and the lamella α extending from this SPP seem to inhibit further blocky α growth; and pinning due to long straight grain boundaries that are lower energy and stable compared to the surrounding area.
- The size and morphology of the blocky α grain growth within the weld region can be predicted from both the region (e.g. HAZ, or near the weld centreline) and from the initial microstructure by reconstructing the parent β microstructure.
- The initial parent microstructure may not be recreated exactly in the post heat treatment microstructure as grain growth can occur at multiple points along the parent β grain boundary, leading to subdivision and/or growth into adjacent prior-β domains.

## 6 Data Availability

The data has been made available via Zenodo [10.5281/zenodo.10689658].

## 7 Acknowledgements

We acknowledge helpful contributions and discussions with Grace Lusby, Mike Rogers, Chris Gourlay, and Fionn Dunne. We acknowledge funding from EPSRC EP/S01702X/1.

## 8 CRediT author statement

Ruth Birch - conceptualization, methodology, formal analysis, investigation, data curation, writing – original draft, visualization. Ben Britton – conceptualization, methodology, resources, writing – review & editing, supervision, project administration, funding acquisition.

1970.

## 10 Supplementary Data

### 10.1 Bulk Microstructure

Bulk microstructure lath measurements pre and post head treatment for the *ex situ* experiments are shown in Table 1.

*Table 1 - α lath dimensions*[1]

| α laths | Pre-HT | | Post-HT | |
|---|---|---|---|---|
| | Map 1 | Map 4 | Map 1 | Map 4 |
| **Mean length (μm)** | 9.4 | 21.9 | 9.5 | 20.8 |
| **Mean width (μm)** | 3.2 | 7.4 | 3.2 | 7.1 |

### 10.2 Neighbourhood prior β orientations

Figure 7 shows a blocky α grain ($\alpha_1$) that has grown across two prior-β domains, as seen in the orientation maps and the reconstructed β maps for pre-HT and post-HT in this example. From the post-HT reconstructed β map, we see the presence of a new parent β orientation for $\alpha_1$ (Figure 7 D), suggesting that this orientation originates from outside the original region (i.e. not from $\beta_1$ or $\beta_2$). The pole figure analysis in Figure 7 also confirms that the blocky-α orientation did not originate in the spatial area ($\beta_1$ or $\beta_2$ domains) that it now fills – there are no overlapping points between $\alpha_1$ and the α grain orientations from the spatial regions $\beta_1$ and $\beta_2$. Analysis of other grains in the map (not explicitly shown here) reveals that this is consistent across the weld.

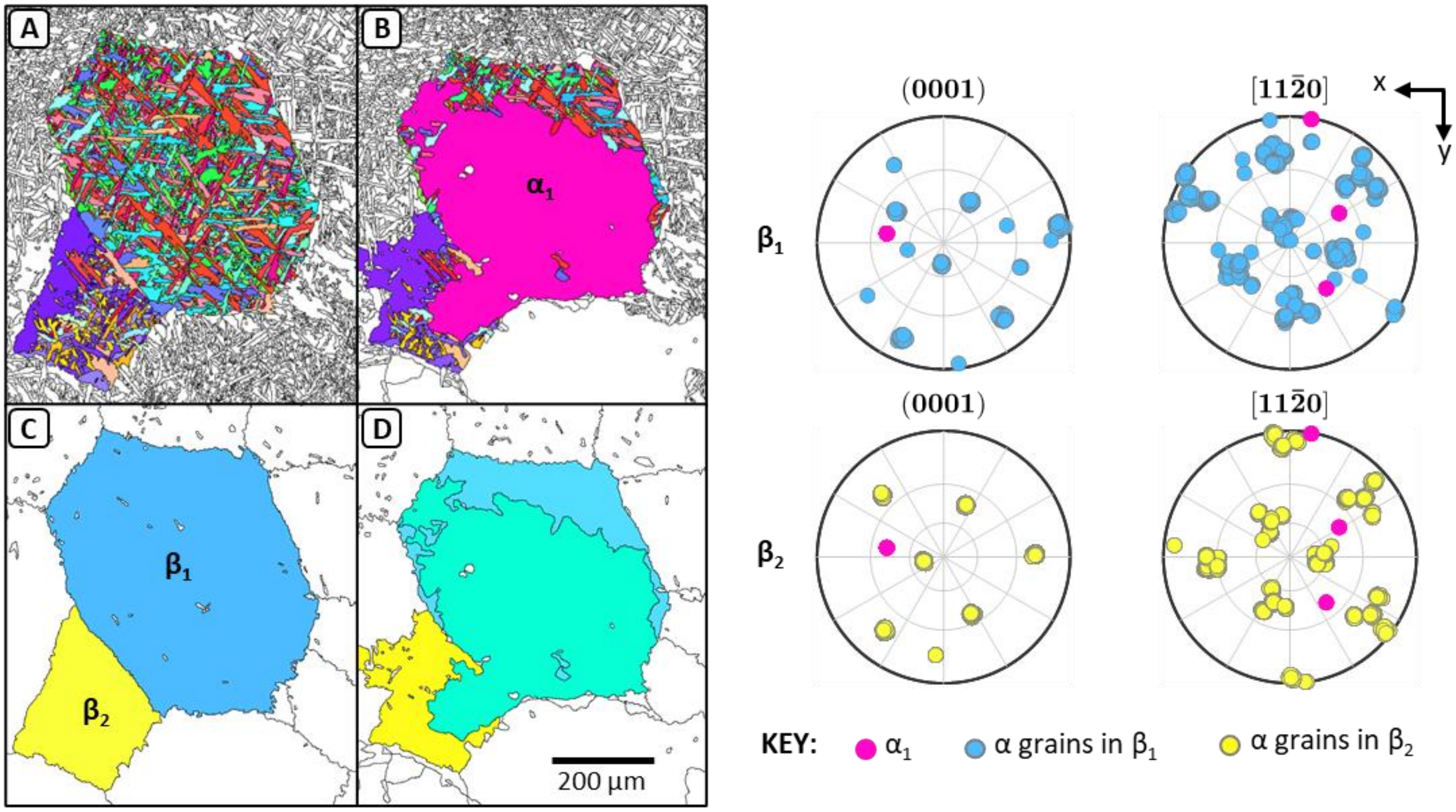

*Figure 7 - Parent β vs. post-HT orientation testing. LEFT: orientation maps for (A) pre-HT α; (B) post-HT α; (C) pre-HT reconstructed β; (D) post-HT reconstructed β using IPFZ colouring. RIGHT: pole figures for the pre-HT α grains contained within spatial regions $\beta_1$ (blue) and $\beta_2$ (yellow) from (C), along with the growing α grain ($\alpha_1$) orientation (smaller pink dots), for $(0001)_\alpha$ and $<11\text{-}20>_\alpha$.*

[1] Map 4 laths: Edges removed; Map 1 laths: Edges and large grains (grainsize>5000 pixels) removed.

Full analysis of the α-variants associated with each of the neighbouring prior-β grains to determine the origin of the growing α grain is shown in Figure 8. Overlapping points ($\alpha_1$ & β) on the pole figures indicates potential β domains from which $\alpha_1$ may have originated, in this case $\beta_5$ or $\beta_{10}$, as indicated by the dashed line and arrows in Figure 8.

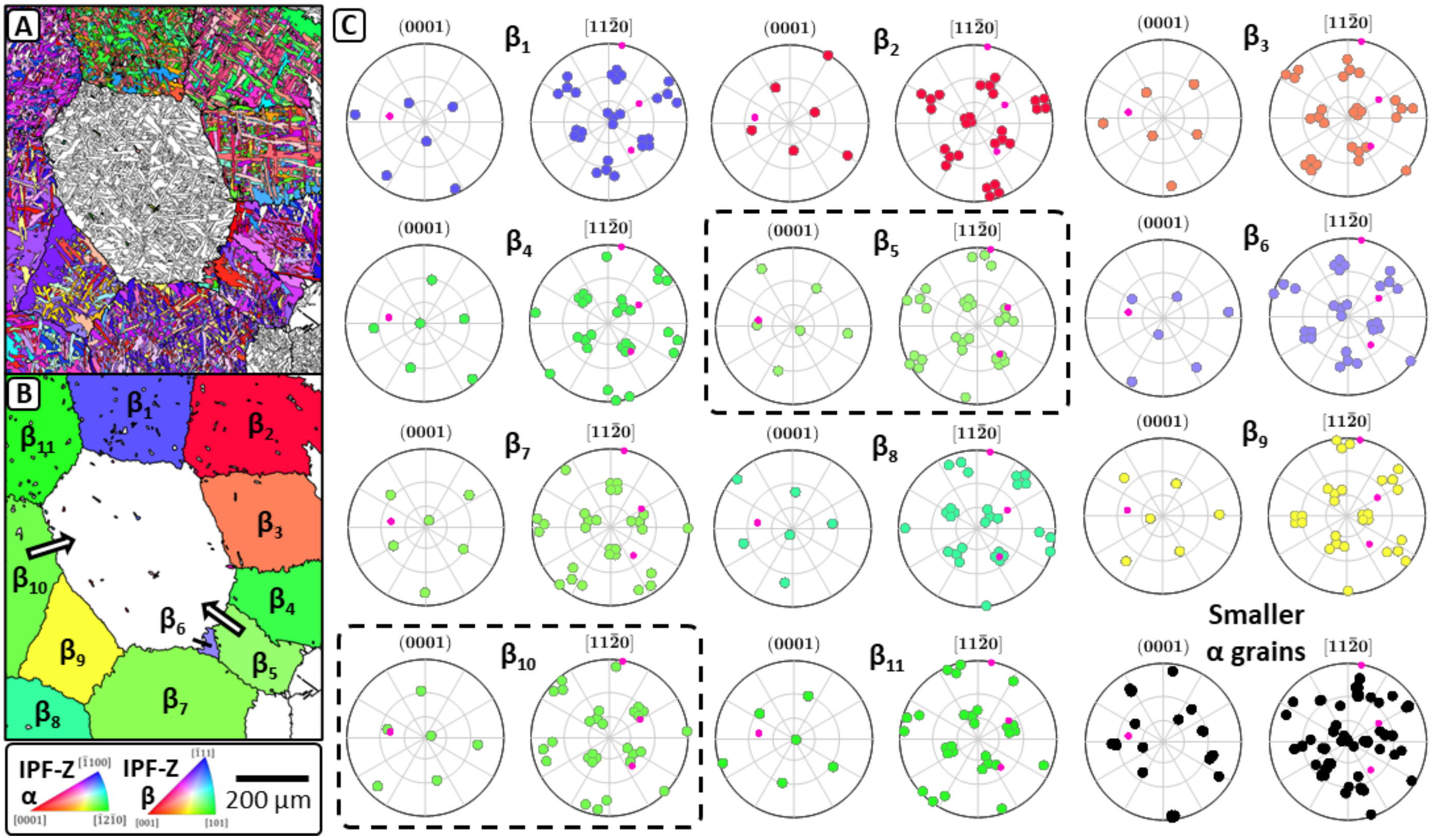

*Figure 8 - Neighbouring grain orientation testing. Neighbouring grains to the parent grain region in which the blocky α grain grows, shown as (A) α grain orientations using IPFZ colouring and (B) parent β grain orientations using IPFZ colouring; (C) pole figures (PF) showing (0001) and <11-20> for the growing α grain orientation (smaller pink dots) and the 12 potential α variant orientations for each of the parent β orientations numbered in (B), shown as the larger coloured dots (IPFZ β colouring). The exception is the last plot which contains the smaller α grains across this region that are not grouped with the larger numbered β grains (black dots). The dashed black rectangle outlines two potential β grains that the growing α grain ($\alpha_1$) may have originated from.*

## 10.3 α variant analysis

Full α variant analysis for the region used in the grain growth model example is shown in Figure 9.

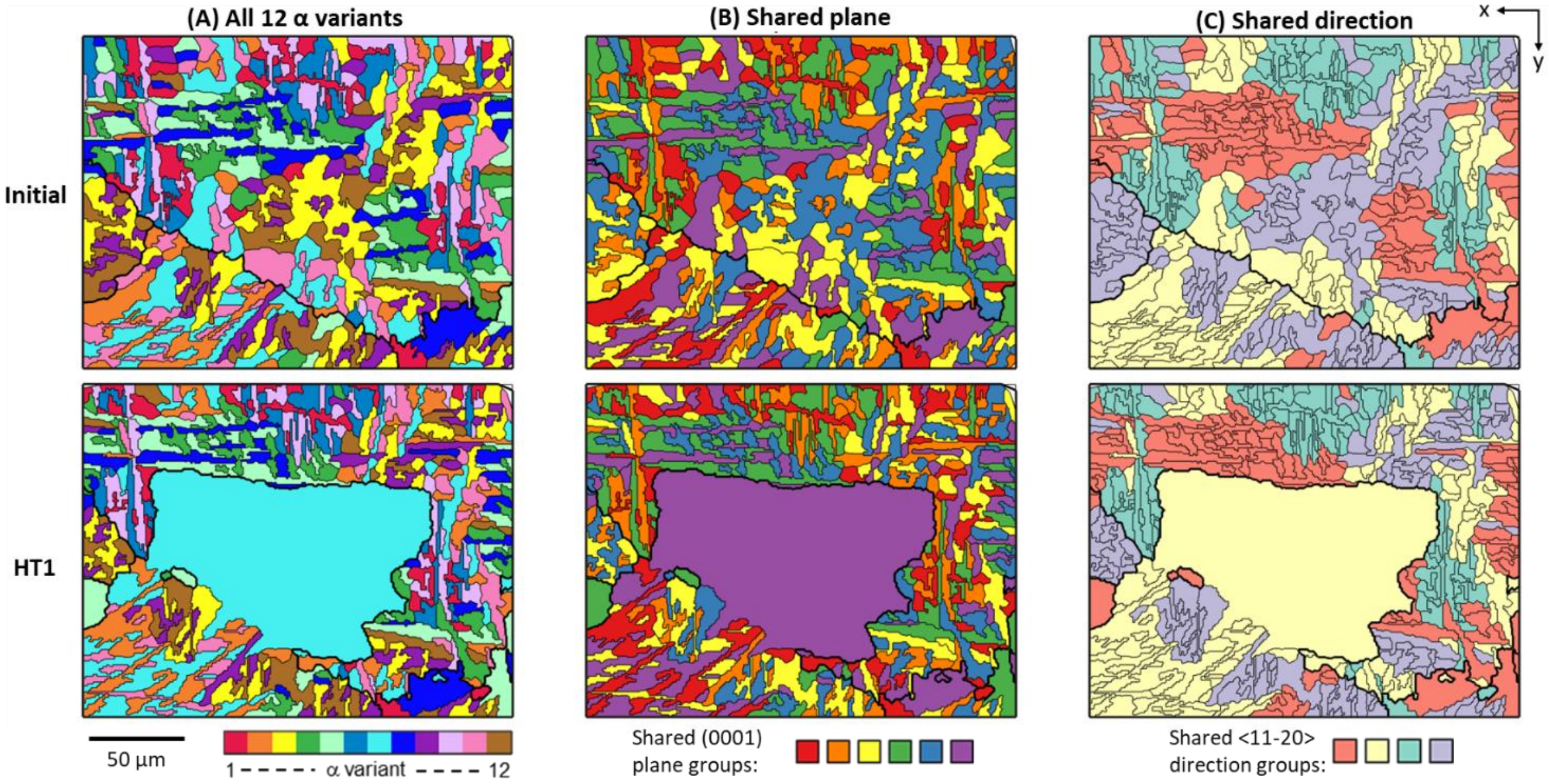

*Figure 9 - α variant analysis for experiment 3 example. Initial and heat treatment 1 (HT1) grain maps for a grain growth example, using α variant colouring. (A) all 12 α variants; (B) coloured by shared (0001) plane (bright colours); (C) coloured by shared <11-20> direction (pastel colours).*

### 10.4 Evaluation of 'cross points'

Looking at the grain orientations emanating from these points, there is often a variety of α variants, with few adjacent grains having shared planes or directions, which may explain the limited growth around these. Three examples are shown in Figure 10, where the maps are coloured by variant number and the prisms show the orientations (with shared planes/directions indicated by shared colours on the prisms).

As can be seen, in some cases, e.g. Area 2 in Figure 10, there are no shared planes or directions between the variants, whereas in other cases, e.g. Area 3 in Figure 10, there are shared directions between adjacent α grains like the purple and blue grains (α variant no. 6 & 10) which share a common direction. In microstructures with more advanced grain growth, these points are not immediately obvious. It is possible that they can be consumed from certain directions using the shared planes/directions. Where there are multiple blocky grains growing, these regions may be potential sites where the growing α grains touch and growth is halted, but further investigation would be required to confirm this.

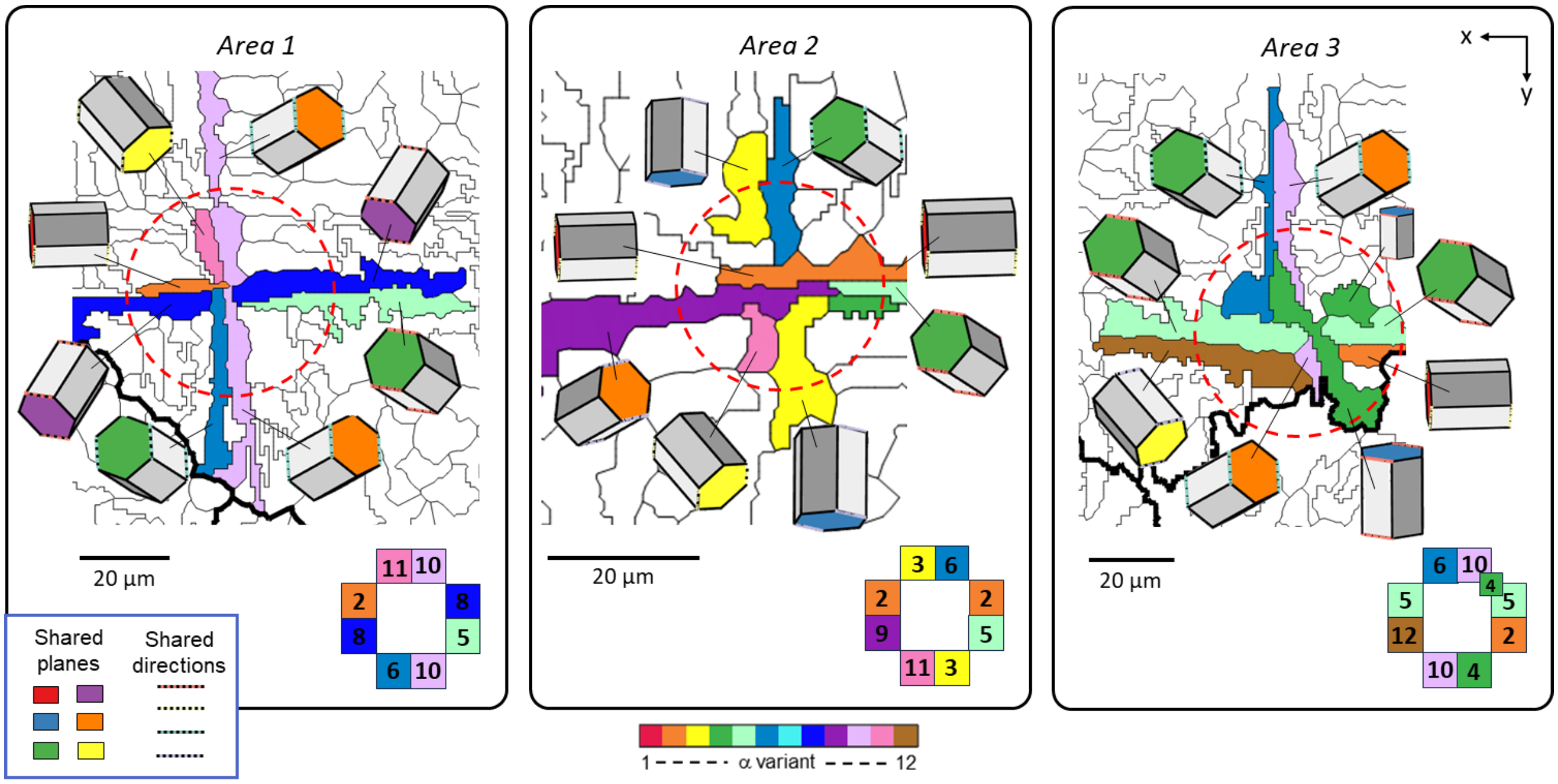

*Figure 10 - 'cross' point α orientation examples. Three areas where 'cross' points exist in an experiment 3 example (from map at Figure 9). α grains coloured by variant number (number also shown in in the adjacent schematics), with orientations indicated by the prisms. Note: prism end colours indicate shared planes, whilst the dashed pastel edge colours indicate shared directions.*